\documentclass[prb,twocolumn,aps,superscriptaddress]{revtex4}
\usepackage{graphicx}
\usepackage{dcolumn}
\usepackage{bm}
\usepackage[dvipsnames]{xcolor}
\usepackage{amssymb,amsmath}
\usepackage{bbold}
\usepackage{ulem}
\usepackage{placeins}

\begin{document}

\title{
Transport signatures of a quantum spin Hall - chiral topological superconductor junction
}

\author{E.G. Novik}
\affiliation{Institute of Theoretical Physics, Technische Universit\"at Dresden, 01062 Dresden, Germany}
\affiliation{Institute of Theoretical Physics and Astrophysics, University of W\"urzburg, D-97074 W\"urzburg, Germany}
\affiliation{Dresden High Magnetic Field Laboratory (HLD-EMFL), Helmholtz-Zentrum Dresden-Rossendorf, 01328 Dresden, Germany}
\author{B. Trauzettel}
\affiliation{Institute of Theoretical Physics and Astrophysics, University of W\"urzburg, D-97074 W\"urzburg, Germany}
\affiliation{W\"urzburg-Dresden Cluster of Excellence ct.qmat, Germany}
\author{P. Recher}
\affiliation{Institute for Mathematical Physics, TU Braunschweig, D-38106 Braunschweig, Germany}
\affiliation{Laboratory for Emerging Nanometrology Braunschweig, D-38106 Braunschweig, Germany}
\begin{abstract}
We investigate transport through a normal-superconductor (NS) junction made from a quantum spin Hall (QSH) system  with helical edge states and a two-dimensional (2D) chiral topological superconductor (TSC) having a chiral Majorana edge mode. We employ a two-dimensional extended four-band model for HgTe-based quantum wells in a magnetic (Zeeman) field and subject to s-wave superconductivity. We show using the Bogoliubov-de Gennes scattering formalism that this structure provides a striking transport signal of a 2D TSC. As a function of the sample width (or Fermi energy) the conductance resonances go through a sequence of $2e^2/h$ (non-trivial phase) and $4e^2/h$ plateaux (trivial phase) which fall within the region of a non-zero Chern number (2D limit) as the sample width becomes large.
These signatures are a manifestation of the topological nature of the QSH effect and the TSC.
\end{abstract}

\date{\today}
\maketitle

\section{Introduction}

The entrance of topology in characterizing the features of materials is a rather recent event \cite{Klitzing80, Laughlin81, Thouless82, Kohmoto85, Berry84, Haldane88, Volovik03, Kane05, Bernevig2006, Koenig07, Hasan10, Qi11}. After the discovery of the quantum Hall effect (QHE) \cite{Klitzing80} and its theoretical description in terms of a topological Chern number relating a bulk property to the existence of chiral edge channels \cite{Thouless82, Kohmoto85}, complementary effects in two-dimensional materials were predicted and discovered, like the quantum spin Hall (QSH) effect \cite{Kane05, Bernevig2006, Koenig07}, possessing helical edge states, the quantum anomalous Hall effect (QAH) \cite{Liu16}, exhibiting chiral edge states, and topological superconductors (TSC) with Majorana edge states \cite{Hasan10, Qi11}. 
\begin{figure}
\includegraphics[width=8.7cm]{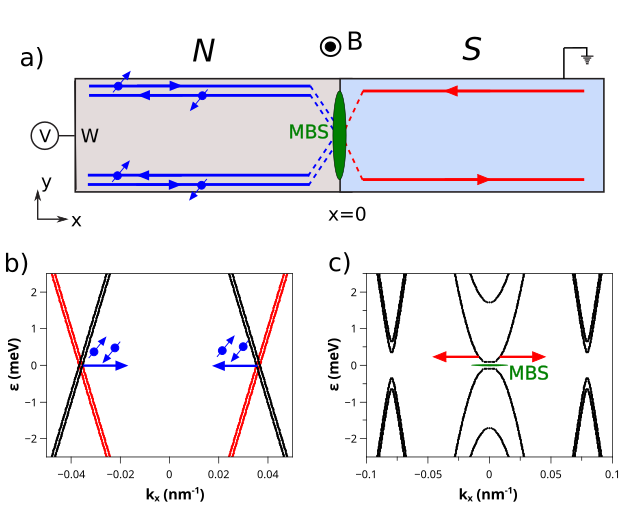}
\caption{a) Scheme of an inverted HgTe QW based NS-junction with width $W$. Superconductivity is induced in the S-region via the proximity effect with a bulk s-wave superconductor ($x >0$). The helical edge channels (blue) are present on the N-side with the Fermi level in the bulk gap. A bias voltage $V$ is applied to the N-contact. In the presence of a magnetic field B, the S-side becomes a TSC with a chiral Majorana edge channel (red). 
b) Dispersion relation of the BdG-spectrum for  electrons (black) and holes (red) for the N-side. Only helical edge states appear within the bulk gap of the QSH insulator. The arrows denote the propagation- and spin-directions of electrons, respectively. c) For $|\Delta|>0$, the QW is turned into a TSC with chiral Majorana edge modes (propagating along the red arrows). The minigap around $k=0$ is due to the mode quantization in the ribbon geometry and can be opened and closed by tuning the magnetic field changing the topological character of the TSC ribbon geometry. In the topologically non-trivial phases a single MBS at $\varepsilon=0$ appears
at the NS-interface, whereas the helical edge channels become gapped out by $\Delta$. We choose $\Delta_E=1.5$ meV, $\Delta_H=0$, $\alpha=0$, $C_{N}=0$, $C_{S}=9.7$ meV, $M=-10$ meV, $E_F=0$, $W=250$ nm.}   
\label{scheme_NS}
\end{figure}

Chiral TSCs recently got considerable attention theoretically \cite{Chung2011,Wang2015, Lian2018, Osca2018} and experimentally \cite{He17} in a QAH-TSC-QAH hybrid system showing evidence of a distinct $e^2/2h$ conductance step. This signature was propagated as an indication of a chiral Majorana edge channel at the boundary of the TSC region. However, subsequent theoretical works \cite{Ji18,Huang18} put forward alternative explanations not related to the existence of a chiral Majorana edge mode. Furthermore, the QAH-TSC system was proposed as a platform for non-Abelian braiding \cite{Lian2017, Beenakker2018}. It is therefore of utmost importance to find additional means to probe a chiral TSC. In the seminal work by Law {\it et al}. \cite{Law09}, the signature of a chiral Majorana edge mode was proposed---in a closed system with finite mode quantization---via $2e^2/h$ tunneling resonances reflecting the Majorana nature of the chiral edge mode. Similarly, Majorana bound states (MBSs) at the ends of proximitized topological semiconducting quantum wires \cite{Mourik12, Das12, Deng12, Lee13, Albrecht16, Zhang18} and in chains of magnetic adatoms on superconductors with spin-orbit coupling \cite{Perge14, Pawlak16} have been probed by tunneling experiments.

We propose a new system presented by an NS-junction composed of a QSH insulator (N-side) with helical edge channels and a TSC (S-side) with a chiral Majorana edge channel in the same material system. In a ribbon geometry using an extended Bernevig-Hughes-Zhang (BHZ) model for an inverted HgTe quantum well (QW) in proximity to an s-wave superconductor and in the presence of a Zeeman field \cite{Weithofer2013, Reuther2013}, we show that the emergence of the chiral TSC is represented by regions of conductance quantization at zero energy being either $4e^2/h$ in a trivial phase of the TSC, reflecting the two spatially separated helical edge channels acting as sources of perfect Andreev reflection, or $2e^2/h$ being the indication of a non-trivial phase for the TSC. The latter phase can be traced back to the existence of a MBS at the NS-interface coupling to a single but spatially separated spinful channel composed of the two helical edge states in the N-region (see Fig.~\ref{scheme_NS}). The non-trivial phases are identified by the parity of the number of bands crossing the Fermi energy in the S-region, reminiscent of a multichannel topological quantum wire \cite{Wimmer2010, Potter2010, Lutchyn2011, Shen2011, Kells2012, Reuther2013}. We stress that the existence of these two distinguishing quantized conductance values is independent of the geometrical details of the setup like the sample width and other imperfections --- owing to the unique setup with two helical channels in the QSH phase acting as detectors of the TSC. These features occur only inside the 2D topological non-trivial phase (Chern number $\pm 1$) when the sample width largely exceeds the extent of the chiral Majorana edge channels which is on the order of the superconducting coherence length $\zeta$. This is therefore a decisive signature of a 2D TSC. We contrast our calculations with the HgTe QW in the non-inverted (without helical edge states) regime where in general multiple channels exist in the N-region and the above mentioned quantization becomes blurred by non-generic and non-quantized conductance values. 

The paper is organized as follows. In Sec. II, we describe the BHZ model for HgTe-based QWs including the effects of Rashba and Dresselhaus spin-orbit interaction, induced s-wave superconductivity and a Zeeman field. In Sec. III, we present the transport properties of NS junctions based on a QSH insulator and a chiral TSC. In Sec. IV, we explain why the appearance of a chiral Majorana edge mode in the TSC expresses itself in quantized $2e^2/h$ conductance plateaux at zero voltage when coupled to helical edge states of a QSH insulator in a strip geometry, whereas a trivial phase of the TSC is characterized by the conductance value of $4e^2/h$. In Sec. V, we predict and demonstrate that the conductance quantization in the topologically non-trivial TSC regime towards higher excitation energies is possible in NN'S junctions. In Sec. VI, we compare our calculations with the case of a HgTe QW in the non-inverted regime. Finally, in Sec. VII, we provide concluding remarks. Additional information is placed in Appendices, as follows. Appendix A provides a detailed description of the model used for the calculations. Appendix B contains details concerning the band structure of the N and S-sides of the QSH-superconductor junction as well as transport properties of the NS junctions based on the inverted and non-inverted HgTe QWs.

\section{Model}

We model the NS junction in a HgTe-based QW by the Bogoliubov-de Gennes (BdG) formalism based on the BHZ model \cite{Bernevig2006} including the effects of Rashba and Dresselhaus spin-orbit interaction \cite{Rothe2010,Zhang2010,Liu2008}. Additionally, we consider a Zeeman field \cite{Koenig2008} and incorporate superconductivity by the proximity effect with an s-wave bulk superconductor. The BdG Hamiltonian for the NS-hybrid structure is then written as $H_{\rm NS}=\int d^{2} r \Psi^{\dagger}({\bm r}){\cal H} \Psi({\bm r})/2$ with
\begin{equation}
\label{BdG1}
{\cal H}=\left(\begin{array}{cc} {\cal H}_e-E_F  &  \Delta  \\ \Delta^{*} & E_F-{\cal H}_h\end{array}\right)
\end{equation}
where ${\cal H}_h={\cal T}{\cal H}_e{\cal T}^{-1}$ is the Hamiltonian for holes with ${\cal T}=is_y \sigma_0 {\cal C}$ the time-reversal operator. $\Delta$ is the induced s-wave pairing potential and $E_F$ is the Fermi energy. We decompose ${\cal H}_e={\cal H}_0+{\cal H}_R+{\cal H}_D +{\cal H}_Z$ in the basis ($|E+\rangle,|H+\rangle,|E-\rangle,|H-\rangle$) \cite{Bernevig2006,basis}, where $E$ ($H$) denotes the electron (heavy hole) subband (SB) and $+$($-$) stands for spin up (down). The bare BHZ Hamiltonian reads ${\cal H}_0=A({\hat k}_x s_z\sigma_x-{\hat k}_y\sigma_y)+ \xi({\hat  k})+M({\hat k})\sigma_z$  with $\xi({\hat k})=C-D{\hat {\bm k}}^2$ and $M({\hat k})=M-B{\hat {\bm k}}^2$,  the Rashba spin-orbit term is ${\cal H}_R=\alpha({\hat k}_ys_x-{\hat k}_xs_y)(\sigma_0+\sigma_z)/2$, and the Dresselhaus spin-orbit term becomes ${\cal H}_D=\delta_0s_y \sigma_y+\delta_e({\hat k}_x s_x-{\hat k}_y s_y)(\sigma_0+\sigma_z)/2+\delta_h({\hat k}_x s_x+{\hat k}_y s_y)(\sigma_0-\sigma_z)/2$ (see Refs.~\onlinecite{Liu2008,Pikulin2014}). The Rashba spin-orbit coupling strength $\alpha$ is tunable by an electric field \cite{Rothe2010}, BIA parameters $\delta_e$, $\delta_h$, $\delta_0$ are material specific \cite{param2} and $A,B,C,D,M$ are band structure parameters \cite{param1}, where the sign of $M$ distinguishes the inverted ($M<0$) regime with helical edge states from the non-inverted (trivial) regime. We also consider the effect of a Zeeman field perpendicular to the plane of the QW \cite{Koenig2008} (see Fig.~\ref{scheme_NS}) described by ${\cal H}_Z=s_z(B_+ +B_-\sigma_z)$ with $B_{\pm}=(\Delta_E\pm\Delta_H)/2$ where $\Delta_E$ and $\Delta_H$ are the Zeeman energies of the $E$ and $H$ bands, respectively. The Pauli matrices $s_i$ and $\sigma_i$ act on spin $(\pm)$ and orbital $(E,H)$ degrees of freedom, respectively and $\sigma_0$ denotes the $2\times2$ identity matrix and ${\hat {\bm k}}=-i\hbar\nabla_{\bm r}$.
       
\section{Transport properties of a QSH-chiral TSC junction}

We consider transport in $x$-direction of an NS structure and assume hard-wall boundary conditions in $y$-direction (see Fig.~\ref{scheme_NS}). The normal (N) region ($x<0$) has $\Delta=0$ and $C=C_N$ whereas the superconducting (S) region ($x>0$) has $\Delta=\Delta_0 e^{i\phi}$ (see Ref.~\onlinecite{param3}) and $C=C_S$. We use a generalized wave-matching method \cite{Zhang2010,Reinthaler2012} in order to solve the Andreev scattering problem for an incoming normal electron with a given excitation energy $\varepsilon$. The corresponding scattering states $\Phi({\bm r})$ solve the BdG equation ${\cal H}\Phi({\bm r})=\varepsilon\Phi({\bm r})$. More details on the approach are given in the Appendix A. 
     
On the N-side of the junction, $C_N$ is chosen such that the Fermi level lies in the bulk gap of the inverted QW and transport proceeds via the helical edge states (see Fig.~\ref{scheme_NS}b)). 
      On the S-side, we tune the Fermi level via $C_S$ to lie in the vicinity of the valence band maximum ($C_S \approx -M$) where the weight of the energy eigenstates is mostly on the $E$-SB for low energies \cite{Weithofer2013}. Since the Zeeman splitting in the $E$-SB is much larger than the one in the $H$-SB \cite{Koenig2008} its influence is maximized. On the contrary, the helical edge states are mainly localized on the $H$-SB \cite{Virtanen2012,Ortiz2016}, so there the Zeeman effect is negligible. For a transparent presentation of our results, we set $\Delta_H=0$ in the following (see also the Appendix B for further discussions).

The subgap conductance at zero temperature can be expressed via the Andreev reflection matrix $r_{he}$ only 
\begin{equation}
G=\frac{2e^2}{h} {\rm Tr}[r_{he}^{\dagger}r_{he}]
\label{conductance1}
\end{equation}
evaluated at a given excitation energy $\varepsilon=eV$ with $V$ the bias voltage applied to the normal contact and $e$ the elementary charge. In Figs.~\ref{conduct_invert}c),e), we present the zero voltage conductance as a function of Fermi energy (or gate voltage) and Zeeman energy $\Delta_E$. 
      
For vanishing or small Zeeman splittings ($\Delta_E<\Delta_0$) where the TSC is in the topologically trivial regime we observe a constant value of $G=4e^2/h$. This is consistent with previous studies \cite{Adroguer2010} at zero magnetic field and in the absence of spin axial symmetry breaking terms  which is a consequence of the spin helicity of the edge states in the QSH insulator. We note that the sample width in our case is finite ($W=250$ nm ($W=1000$ nm) in Fig.~\ref{conduct_invert} a)-d) (e) and f)) leading to a small overlap of the helical edge states near the Dirac point. For a Fermi level away from the Dirac point, this hybridization, however, is effectively suppressed leading to two separate propagating channels with perfect Andreev reflection. 
      
If $\Delta_E > \Delta_0$, the zero bias conductance switches between  $4e^2/h$ and $2e^2/h$ (see Fig.~\ref{conduct_invert} c),e)), depending on the number $N$ of bulk subbands crossing the Fermi level on the S-side in the absence of $\Delta$ (see Figs.~\ref{conduct_invert} a), f)). 
      
\section{Chiral Majorana edge modes}

\begin{center}            
\begin{figure}
\includegraphics[width=14cm]{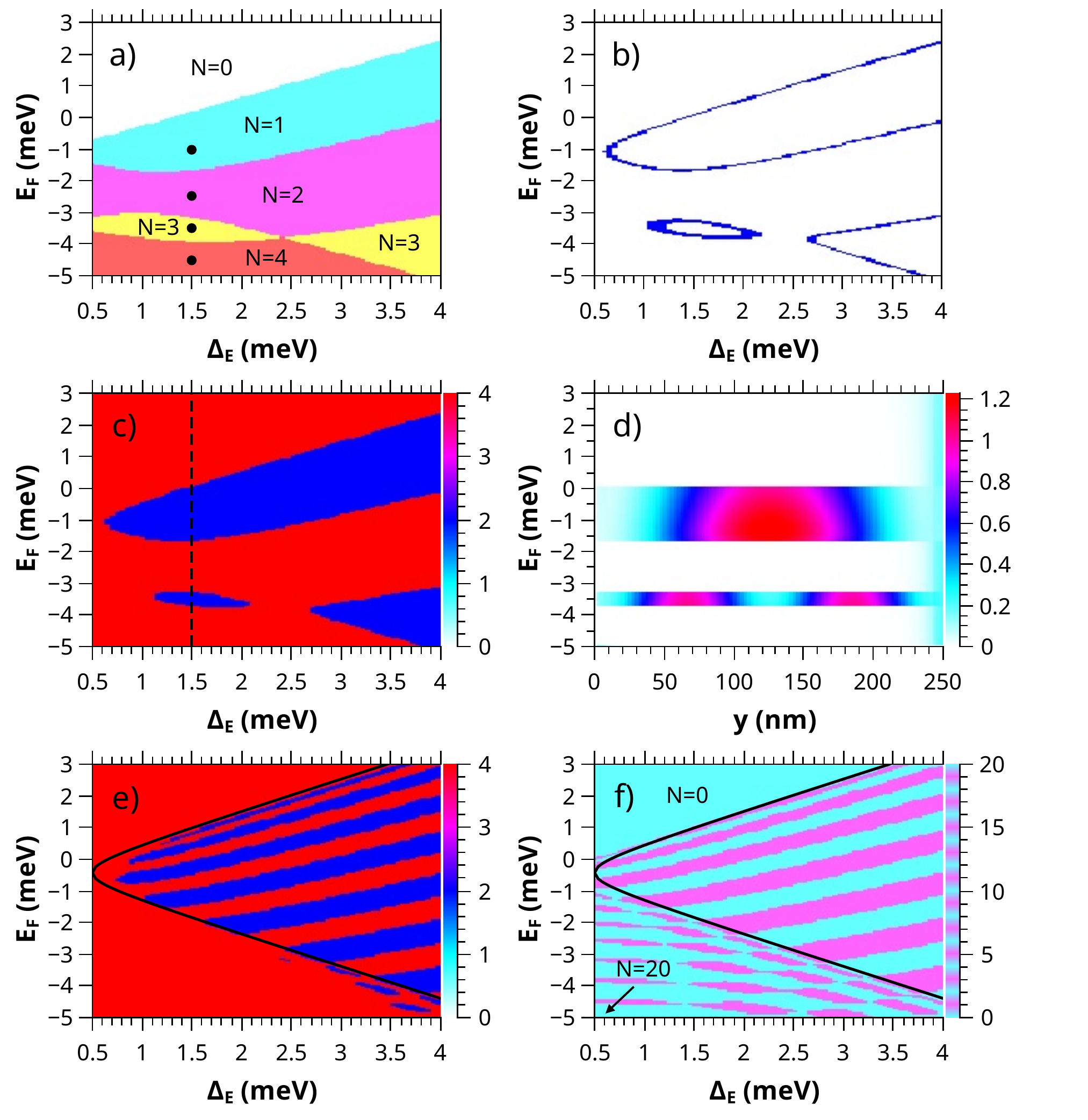}
\caption{a) Diagram representing the number of occupied bulk subbands at the Fermi energy in an inverted QW (edge state subbands are not counted here); b) range (blue regions) where the bulk gap in the superconducting QW is negligibly small (i.e. the gap $\leq0.06$~meV); c) conductance (in units of $e^2/h$) in the NS structure based on an inverted QW; d) probability density at the interface of N- and S-layers. All plots are presented for the QW with a width $W=250$~nm, $M=-10$~meV, $\alpha=0$. Other parameters are $\varepsilon=0$ in a), c) and d), the doping parameter is $C_N=9.7$~meV in a) and $C_S=9.7$~meV in b), $C_N=0$, $C_S=9.7$~meV in c) and d). The Zeeman term is set to 1.5~meV in d), which corresponds to the dashed line in c). Plots e) and f) correspond to c) and a), resp. for $W=1000$ nm, with the black line given by $\Delta _E^2=\Delta _0^2+(C_S+M+\delta _0^2/(2M)-E_F)^2$ in accordance with Ref.~\onlinecite{Weithofer2013}. Regions with odd (even) values of $N$ are shown in magenta (light blue) in f).}
\label{conduct_invert}
\end{figure}
\end{center}

We now explain why the switching from a $4e^2/h$ to a $2e^2/h$ conductance plateau at $V=0$ is a direct qualitative  transport signature of a TSC induced by the presence of Dresselhaus and/or Rashba spin orbit coupling in the HgTe QWs when $\Delta_E > \Delta_0$ (see Refs.~\onlinecite{Weithofer2013, Reuther2013}). The hallmark of 2D TSCs is the appearance of chiral Majorana edge modes. In the ribbon geometry, the two counterpropagating chiral Majorana edge channels at the opposite edges develop a minigap around $\varepsilon=0$ due to the finite width of the ribbon (see Fig.~\ref{scheme_NS}c)). The closing of these minigaps in the BdG dispersion relation [see Fig.~\ref{conduct_invert}b)] happens at the topological phase transition between trivial and non-trivial superconducting phases, each of them characterized by even and odd values of $N$, respectively [see Fig.~\ref{conduct_invert}a) and the Appendix B for examples of dispersion relations]. Note that the tails of the helical edge states at higher $k$-values, are gapped by the superconductor. 
 If $N$ is odd the S-ribbon is topologically non-trivial \cite{Reuther2013} with an associated MBS at exactly zero energy ($\varepsilon=0$) located at the boundary to the normal side of the NS-junction at $x=0$. The helical edge channels on the N-side couple to the MBS such that the eigenvalues of  $r_{he}^{\dagger}r_{he}$ become non-degenerate and equal to 1 and 0 which is a sign of the non-trivial chiral TSC \cite{Beri2009, Akhmerov2011, Wimmer2011, Fulga2011}. A non-degenerate unit Andreev reflection eigenvalue results here in a quantized conductance of $2e^2/h$. On the contrary, if $N$ is even then $G=4e^2/h$ and the TSC is trivial.

\onecolumngrid
\twocolumngrid

In the limit of large $W$ (Fig.~\ref{conduct_invert} e) and f)), conductance plateaux with $G=2e^2/h$ (non-trivial TSC) become dense and fall into the region of a non-trivial Chern number $\mathcal{C}=-1$ (to the right of the full line in Fig.~\ref{conduct_invert} e)) of the 2D TSC \cite{Weithofer2013} with a chiral Majorana edge mode. Outside this region, $\mathcal{C}=0$ and the conductance of the NS-junction is $G=4e^2/h$ (and the TSC trivial), independent on whether the parity of occupied bulk subbands at the Fermi energy is odd or even (Fig.~\ref{conduct_invert} f)). In this sense this setup allows to probe the {\it 2D Chern number of the TSC}.

Information on the MBS is also contained in the scattering states of our NS-system. We plot the probability density of these scattering states as obtained in Appendix A in Fig.~\ref{prob_density}b) for $W=250$ nm. Contrary, in the trivial case, these bound states are absent and only the incoming electron-like and the reflected hole-like helical edge states are visible (see Fig.~\ref{prob_density}a)). We also depict the presence of these bound states along the transverse direction as a function of $E_F$ in Fig.~\ref{conduct_invert}d) consistent with the corresponding conductance values in Fig.~\ref{conduct_invert}c) and spectral properties in Figs.~\ref{conduct_invert} a),b). When $W \gg \zeta$ the MBS appears to have the property of an extended state (see Fig.~\ref{prob_density}c) for $W=1000$ nm) which we interpret as a crossover from a wire-like regime to a more 2D TSC regime, consistent with alternative setups \cite{Potter2010, Kells2012}. It clearly shows the onset of a chiral Majorana edge state at the boundary of the S-region.
\begin{figure}
\includegraphics[width=6.5cm]{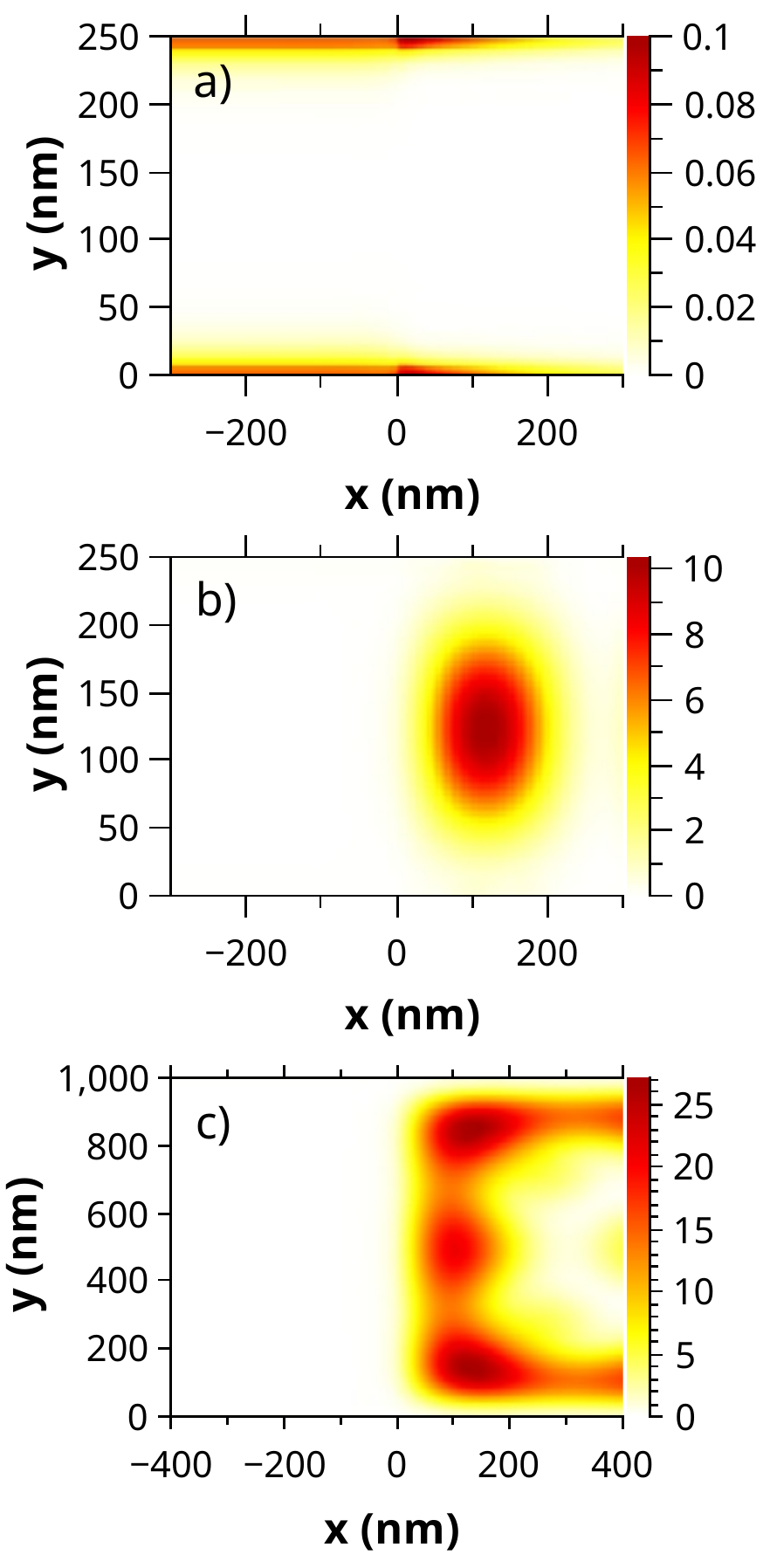}
\caption{Probability density $\vert\Phi(x,y)\vert^2$ of scattering states at $\varepsilon=0$ in the NS structure based on an inverted QW for the topologically trivial case in a) and the topologically non-trivial regime in b) and c). The calculations are presented for $M=-10$~meV, $\alpha=0$ and doping parameters $C_N=0$, $C_S=9.7$~meV. Further parameters are $W=250$~nm, $E_F=0$ in a) and b) and $W=1000$~nm, $E_F=-0.6$~meV in c). The Zeeman term is $\Delta_{E}=0$ in a), $\Delta_{E}=2$~meV in b) and $\Delta_{E}=0.8$~meV in c). Due to the very high probability density of the bound state in the S-region in comparison with that of the edge states the latter are not visible in b) and c).}  
\label{prob_density}
\end{figure}

\section{NN'S-junction}

\begin{figure}
\includegraphics[width=8.5cm]{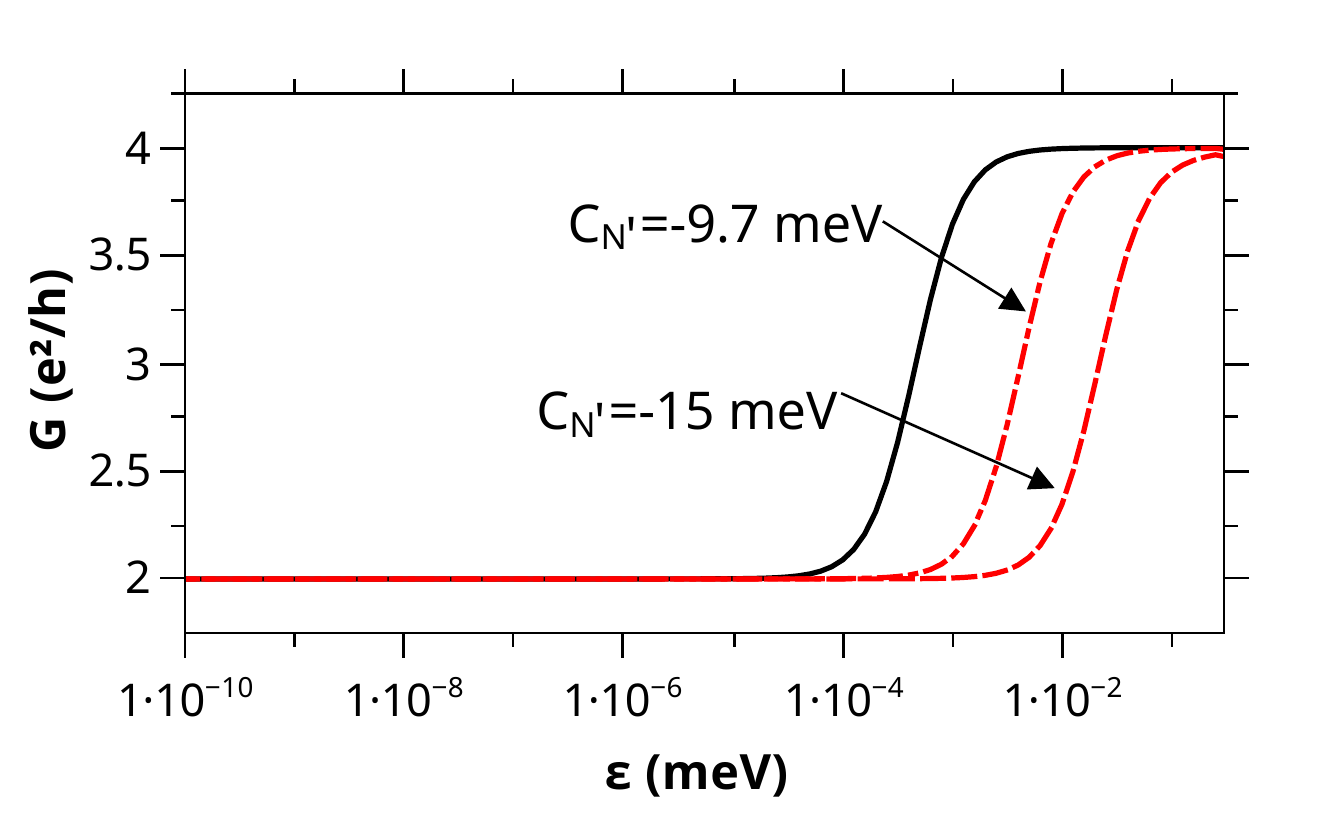}
\caption{Conductance of the NS (solid black line) and NN'S (dot-dashed and dashed red lines) structure as a function of the excitation energy. The calculations are presented for the following parameters: $W=250$~nm, $M=-10$~meV, $E_F=-1$~meV, $\Delta_{E}=1$~meV, $\alpha=0$; the doping parameter is $C_N=0$ and $C_S=9.7$~meV. Thickness of the N'-layer $L=100$~nm.} 
\label{conduct_energy}
\end{figure}

The helical edge states do only couple weakly to the MBS. This is expressed via a sharp resonance as a function of $\varepsilon$ which we display in Fig.~\ref{conduct_energy} (full line). To observe the conductance quantization due to such resonances the energy broadening should exceed the temperature. In this sense, the resonance width sets the temperature at which the experiment should be performed. We observe that the overlap between the MBS and the helical edge states can be enhanced by an intermediate N' layer of length $L$ that has a different $C_N$ parameter (see Fig.~\ref{conduct_energy}). The states in this N' layer couple more efficiently to the MBS which allows to observe the conductance quantization towards higher excitation energies $\varepsilon$. By increasing the Fermi level into the bulk states in the N' layer, we observe a two-orders-of-magnitude increase in $\varepsilon$ at which the resonance is still seen (dashed line). This should make it feasible to observe the MBS resonances in current state of the art experiments in HgTe QWs.

\section{Non-inverted $\textbf{HgTe}$ QWs}

In the non-inverted QWs, the TSC phases are still possible (the sign of $M$ does not influence the topology of the TSC) but the normal lead ceases to have helical edge states. Similar to the case of the QSH insulator (see Fig.~\ref{conduct_invert}) the S-phase is related to the number $N$ of the occupied bulk SBs in the absence of the pairing potential. But in contrast to the QSHI-S junction, the conductance takes on quantized values $G=(2e^2/h)n$, where $n$ is odd (even) for the topologically non-trivial (trivial) S-phase. Moreover, the conductance quantization in the trivial phase is not protected due to imperfect Andreev reflection in the absence of helical edge modes in non-inverted QWs (see Appendix B for more details). Similar behavior has been reported for multichannel semiconducting nanowires in proximity to an s-wave S \cite{Wimmer2011}. This shows that the QSH insulator-TSC junction has rather unique and stable quantized conductance features not seen in other systems which rely on the combination of two topological phases --- the QSH insulator and the chiral TSC.

\section{Conclusion}
We have shown how a decisive signature of a chiral TSC can be observed in an NS-junction based on a QSH insulator and a TSC made of the same material in contact to an s-wave bulk superconductor and subjected to a magnetic field. Using the extended two-dimensional BHZ model (including axial spin symmetry breaking terms, induced s-wave superconductivity and a Zeeman field) in a ribbon geometry -- which takes into account bulk as well as edge states -- we show that the signature of a chiral Majorana edge mode in the TSC part expresses itself in quantized $2e^2/h$ conductance plateaux at zero voltage. These resonances can be traced back to Majorana bound states (MBS) appearing at the NS interface in this ribbon geometry. Moreover, a gate voltage can be used to tune the topological phase of the TSC, resulting in clearly separated quantized conductance plateaux of $2e^2/h$ (topologically non-trivial with MBS) and $4e^2/h$ (topologically trivial without MBS).  For large ribbon width, the topological conductance quantization of $2e^2/h$ can only be seen in the parameter range with non-zero Chern number of the 2D TSC. Hence, we suggest that our proposal constitutes a new way to detect a chiral TSC in two dimensions via transport measurements without the necessity of fine tuning of parameters.

We thank R.W. Reinthaler for helpful discussions. We acknowledge financial support by DFG Grants No. AS 327/5-1 and RE 2978/7-1, the SFB 1143 (project-id 247310070), the SFB 1170 ToCoTronics, the W\"urzburg-Dresden Cluster of Excellence ct.qmat (EXC 2147, project-id 39085490) and the DFG Excellence Strategy-EXC-2123 QuantumFrontiers-390837967.

\appendix
\section{Detailed description of the model}

We describe NS junctions based on HgTe QWs within the BdG formalism and use the BHZ model for the QW \cite{Bernevig2006} including the effects of Rashba and Dresselhaus spin-orbit interaction as well as the Zeeman splitting induced by a magnetic field \cite{Rothe2010,Zhang2010,Liu2008,Koenig2008}. 

We solve the BdG equations ${\cal H}\Phi(\bm{r})=\varepsilon\Phi(\bm{r})$ where the Hamiltonian is given explicitly by 
\begin{widetext}
\begin{eqnarray}
\label{BdG0}
&&{\cal H}=\nonumber\\
&&\left(\begin{array}{cccccccc} 
\substack{\xi({\hat k})+M({\hat k})\\+\Delta_E-E_F} & A{\hat k}_+ & i\alpha {\hat k}_-+\delta_e {\hat k}_+ & -\delta_0 & \Delta & 0 & 0 & 0\\ 
A{\hat k}_- & \substack{\xi({\hat k})-M({\hat k})\\+\Delta_H-E_F} & \delta_0 & \delta_h {\hat k}_- & 0 & \Delta & 0 & 0\\ 
-i\alpha {\hat k}_++\delta_e {\hat k}_- & \delta_0 & \substack{\xi({\hat k})+M({\hat k})\\-\Delta_E-E_F} & -A{\hat k}_- & 0 & 0 & \Delta & 0\\ 
-\delta_0 & \delta_h {\hat k}_+ & -A{\hat k}_+ & \substack{\xi({\hat k})-M({\hat k})\\-\Delta_H-E_F} & 0 & 0 & 0 & \Delta\\ 
\Delta^* & 0 & 0 & 0 & \substack{E_F-\xi({\hat k})\\-M({\hat k})+\Delta_E} & -A{\hat k}_+ & -i\alpha {\hat k}_--\delta_e {\hat k}_+ & \delta_0  \\ 
0 & \Delta^* & 0 & 0 & -A{\hat k}_- & \substack{E_F-\xi({\hat k})\\+M({\hat k})+\Delta_H} & -\delta_0 & -\delta_h {\hat k}_-\\ 
0 & 0 & \Delta^* & 0 & i\alpha {\hat k}_+-\delta_e {\hat k}_- & -\delta_0 & \substack{E_F-\xi({\hat k})\\-M({\hat k})-\Delta_E} & A{\hat k}_-\\ 
0 & 0 & 0 & \Delta^* & \delta_0 & -\delta_h {\hat k}_+ & A{\hat k}_+ & \substack{E_F-\xi({\hat k})\\+M({\hat k})-\Delta_H} 
\end{array} \right),\nonumber\\		
\end{eqnarray}
\end{widetext}
with $\xi({\hat k})=C-D{\hat {\bm k}}^2$, $M({\hat k})=M-B{\hat {\bm k}}^2$ and ${\hat k}_{\pm}={\hat k}_x{\pm}i{\hat k}_y$. We assume a step-like profile for the pairing potential and the doping parameter in the NS structure, i.e. $\Delta=0$ and $C=C_N$ in the N-region ($x<0$) whereas $\Delta=\Delta_0 e^{i \phi}$ and $C=C_S$ in the S-region ($x\geq0$). The operator hat "$\,{\hat{~}}\,$" implies that ${\bm k}$ should be replaced by $-i\hbar\nabla_{\bf r}$ whenever it acts on the spinor $\Phi(\bm{r})$.

We consider transport in $x$ direction of the NS structure and choose hard-wall boundary conditions in $y$ direction. Following the procedure from Refs.~\onlinecite{Zhang2010,Reinthaler2012}, we expand the wave functions in terms of Fourier modes $\varphi_n(y)=\sqrt{2/W}\sin(n\pi y/W)$:
\begin{equation}
\label{Psi}
\Phi_m(x,y)=e^{i k_x^m x}\sum_{n=1}^{N_{\rm{max}}}a_n^m \varphi_n(y),
\end{equation}
where $m$ is an index for different values of the longitudinal momentum for a given excitation energy $\varepsilon$, the number of transverse modes $N_{\rm{max}}$ is chosen to be large enough to ensure the convergence of the numerical solution, and an eight-component spinor $a_n^m$ and momentum $k_x^m$ are determined by solving the eigenvalue problem:
\begin{eqnarray}
\label{EVP}
\left(\begin{array}{cc} \mathbb{1}  &  \mathbb{0}  \\ \mathbb{0} & (H^{k_x^2})^{-1}\end{array}\right)
\left(\begin{array}{cc} \mathbb{0}  &  \mathbb{1}  \\ H^{\rm{const}}+H^{k_y} & H^{k_x}\end{array}\right)
\left(\begin{array}{c} a^m  \\ a'^m\end{array}\right)  \nonumber\\
=k_x^m\left(\begin{array}{c} a^m  \\ a'^m\end{array}\right).
\end{eqnarray}
The vectors $a^m=(a_1^m,a_2^m,...)^T$, $a'^m=(a'^m_1,a'^m_2,...)^T$, $a'^m_n=k_x^m a_n^m$, and the $8\times8$ sub-matrices in Eq.~(\ref{EVP}) have the form:
\begin{eqnarray}
\label{submatrices}
H_{n1, n2}^{k_x^2}&=&\delta_{n1,n2}[(D+B \sigma_z) \tau_z],\nonumber\\
H_{n1, n2}^{\rm{const}}&=&\delta_{n1,n2}[(C+M\sigma_z-E_F+\delta_0 s_y \sigma_y)\tau_z-\varepsilon \nonumber\\
                       &+&(B_+ +B_-\sigma_z)s_z+(\Delta_+\tau_x+i\Delta_-\tau_y)],\nonumber\\
H_{n1, n2}^{k_x}&=&\delta_{n1,n2}[(A s_z \sigma_x+(-\alpha s_y+\delta_e s_x)(\sigma_0+\sigma_z)/2\nonumber\\
                &+&\delta_h s_x (\sigma_0-\sigma_z)/2)\tau_z],\nonumber\\
H_{n1, n2}^{k_y}&=&[-(D+B \sigma_z) \tau_z]P_{n1,n2}\nonumber\\
                &+&[(-A \sigma_y+ (\alpha s_x-\delta_e s_y)(\sigma_0+\sigma_z)/2\nonumber\\
                &+&\delta_h s_y(\sigma_0-\sigma_z)/2)\tau_z] G_{n1,n2}.
\end{eqnarray}
We define $P_{n1,n2}=(\frac{n_1 \pi}{W})^2 \delta_{n1,n2}$, $G_{n1,n2}=\langle\varphi_{n_1}(y)\vert -i \partial_y\vert\varphi_{n_2}(y)\rangle$, $B_{\pm}=(\Delta_E\pm\Delta_H)/2$, $\Delta_{\pm}=(\Delta\pm\Delta^*)/2$ and the Pauli matrices $s_i$, $\sigma_i$ and $\tau_i$ act on spin $(\pm)$, orbital $(E,H)$ and electron-hole degrees of freedom, respectively and $\sigma_0$ denotes the $2\times2$ identity matrix.  

The wave function in the N-layer ($x< 0$) can be taken in the form:
\begin{eqnarray}
\label{PsiN}
\Phi^N(x,y)&=&\Phi_{N_{Re}}(x,y)+
              \sum_{N_{Le}}r_{N_{Le},N_{Re}}\Phi_{N_{Le}}(x,y) \nonumber\\ &+&
              \sum_{N_{Lh}}r_{N_{Lh},N_{Re}}\Phi_{N_{Lh}}(x,y)
\nonumber\\&+&\sum_{N_{Ev}}r_{N_{Ev},N_{Re}}\Phi_{N_{Ev}}(x,y),
\end{eqnarray}
and includes propagating states, i.e. incoming electrons (with index $N_{Re}$ moving to the right ($R$) along the positiv x-axis, see Fig.~\ref{scheme_NS}) and reflected electrons (with indices $N_{Le}$ moving to the left ($L$)) and holes (with indices $N_{Lh}$ moving to the left ($L$)), respectively, as well as evanescent solutions decaying to the left (with indices $N_{Ev}$). Note that, in general, there are several reflected and evanescent modes for a given incoming mode $N$.

The wave function in the S-region ($x\geq0$) takes the form:
\begin{eqnarray}
\label{PsiS}
\Phi^S(x,y)&=& \sum_{S_{R }}t_{S_{R },N_{Re}}\Phi_{S_{R }}(x,y)
\nonumber\\&+& \sum_{S_{Ev}}t_{S_{Ev},N_{Re}}\Phi_{S_{Ev}}(x,y),
\end{eqnarray}
with the evanescent solutions exponentially decaying for $x\rightarrow \infty$ (with index $S_{Ev}$) and transmitted propagating states (with index $S_{R}$). Like in Refs.~\onlinecite{Zhang2010,Reinthaler2012}, we determine the reflection amplitudes of the electron ($r_{N_{Le},N_{Re}}$) and hole ($r_{N_{Lh},N_{Re}}$) states in the left lead and transmission amplitudes ($t_{S_{R},N_{Re}}$) of the states in the right lead by matching the wave functions $\Phi(x,y)$ and the currents $[\partial_{k_x} {\cal H}] \Phi(x,y)$ at the NS-interface $x=0$. Reflection and transmission amplitudes should be renormalized in order to take into account different current densities for the incident, reflected and transmitted states:
\begin{eqnarray}
\label{Renormalization}
r_{ee}^{N_{Le},N_{Re}}&=&r_{N_{Le},N_{Re}}\sqrt{\dfrac{v_{N_{Le}}}{v_{N_{Re}}}},\nonumber\\
r_{he}^{N_{Lh},N_{Re}}&=&r_{N_{Lh},N_{Re}}\sqrt{\dfrac{v_{N_{Lh}}}{v_{N_{Re}}}}, \\
t_{ee}^{S_{R },N_{Re}}&=&t_{S_{R },N_{Re}}\sqrt{\dfrac{v_{S_{R }}}{v_{N_{Re}}}},\nonumber
\end{eqnarray}
where the velocity $v_m$ is given by
\begin{equation}
\label{velocity}
v_m=\hbar^{-1}\int_0^W dy~\Phi_m^{\dagger}(x,y) [\partial_{k_x} {\cal H}]_{k_x\rightarrow k_x^m} \Phi_m(x,y).
\end{equation}
With this renormalization, the propagating states all carry the same particle current. Here, $r_{ee}^{N_{Le},N_{Re}}$,  $r_{he}^{N_{Lh},N_{Re}}$ are the associated reflection probability amplitudes for an incoming electron in mode $N_{Re}$ into an electron in mode $N_{Le}$ or a hole in mode $N_{Lh}$, respectively, and $t_{ee}^{S_{R },N_{Re}}$ is the probability amplitude for the transmission of the incoming electron into mode $S_{R }$ in S.                                                                                                                                                                                                                                 

The differential conductance of the NS structure can be calculated using the Blonder-Tinkham-Klapwijk formula \cite{Blonder1982_S}
\begin{eqnarray}
\label{BTK}
G&=&\frac{e^2}{h}\int d\varepsilon (-\partial_{\varepsilon} f(\varepsilon-eV))\nonumber\\
&\times & \mathrm{Tr} [\mathbb{1}
-r_{ee}^{\dagger}(\varepsilon)r_{ee}(\varepsilon)
+r_{he}^{\dagger}(\varepsilon)r_{he}(\varepsilon)],
\end{eqnarray}
where $f(\varepsilon-eV)$ is the Fermi distribution function, $V$ is the bias voltage applied to the N-region, $r_{ee}$ and $r_{he}$ are normal and Andreev reflection matrices, respectively. In the regime of zero temperature and subgap transport (i.e. the transmission probability $t_{ee}^{\dagger}(\varepsilon)t_{ee}(\varepsilon)=0$, where $\varepsilon=eV$) Eq.~(\ref{BTK}) takes the form:
\begin{equation}
G=\frac{2e^2}{h} {\rm Tr}[r_{he}^{\dagger}(\varepsilon)r_{he}(\varepsilon)].
\label{conductance}
\end{equation}

A topological quantum number
$
Q=(-1)^\mathcal{M}={\rm{sgn}}(\mathrm{Det}~r) 
$
can be exploited to determine the number of topologically protected (quasi)bound states $\mathcal{M}$ at the end of the TSC in the presence of particle-hole symmetry and in the absence of time-reversal and spin-rotation symmetry \cite{Akhmerov2011, Fulga2011}. $Q$ is characterized by the reflection matrix $r$ which has the property $\mathrm{Det}~r=(-1)^{d_u}$, where $d_u$ is the degeneracy of the unit Andreev reflection eigenvalue \cite{Beri2009}.

\section{Details of the band structure and transport properties}

In this Appendix, we present additional information concerning the band structure of the N and S-sides of the QSH-superconductor junction as well as transport properties of the NS junctions for the case of the inverted and non-inverted HgTe QWs.

\subsection{QSHI-S junctions}

\begin{figure}[ht!]
\includegraphics[width=4.5cm]{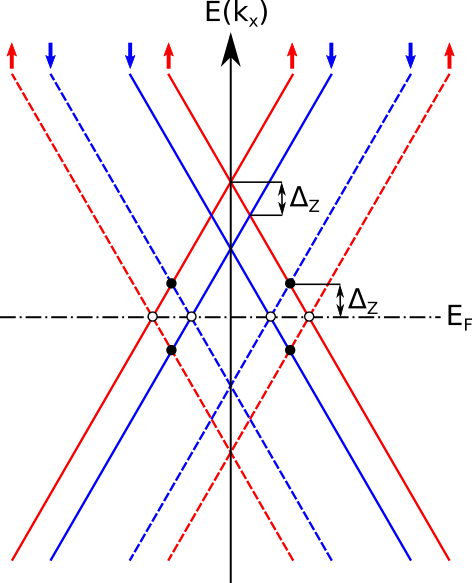}
\caption{Schemes of the electron (solid lines) and hole (dashed lines) edge-state dispersion (in the absence of the pairing potential) in an inverted HgTe QW. Spin-up (spin-down) states are shown by red (blue) lines.} 
\label{edge_states}
\end{figure}

We first consider aspects of the band structure of inverted HgTe QWs including the proximity effect due to an s-wave superconductor. Initially we examine the influence of the Zeeman splitting on the energy dispersion of the edge states.  
In the absence of the pairing potential $\Delta$, an inverted HgTe QW has a band gap in the bulk and helical edge states within this gap \cite{Bernevig2006}. To illustrate the main effects of the influence of the Zeeman term and the induced superconductivity on the helical edge states, the spin-axial breaking Rashba and Dresselhaus terms can be neglected as their influence is rather weak \cite{Virtanen2012, Michetti2012_S} in HgTe QWs for not too strong Rashba coefficients $\alpha$. Therefore, we can use the following equations for the spin-up ($\uparrow$) and the spin-down ($\downarrow$) edge states in a large $W$ limit for energies in the bandgap region \cite{Zhou2008_S}:
\begin{eqnarray}
\label{eq_edge_states}
E_{e\pm}^{  \uparrow}&=&C-\frac{M D}{B}\pm A\sqrt{\frac{B^2-D^2}{B^2}}k_x+\Delta_Z, \nonumber\\ 
E_{e\pm}^{\downarrow}&=&C-\frac{M D}{B}\pm A\sqrt{\frac{B^2-D^2}{B^2}}k_x-\Delta_Z,
\end{eqnarray}
where $\Delta_Z$ is the Zeeman term and we assume that $\Delta_Z=\Delta_{E}=\Delta_{H}$. It should be noted that we choose the position of the Fermi level in N- an S-layers away from the Dirac point, thus we do not suffer from the opening of gaps in the helical edge-state spectrum by finite size effects which are found to be negligibly small for the structures with a width $W\geq 250$~nm (see Fig.~1 in Ref.~\onlinecite{Zhou2008_S}). In the proximity to an s-wave superconductor electron states and their time-reversed partners (i.e. holes with opposite spin orientations) are coupled by the pairing potential $\Delta$. A scheme of the electron and hole edge-state dispersion without their coupling is shown in Fig.~\ref{edge_states}. One can see that the electron and hole bands for the states with the same spin direction, which are not coupled by the pairing potential, cross at $E=E_F$ (crossing points are marked by the open circles in the figure). In contrast, around the energy values $E=E_F\pm\ \Delta_Z$ (black circles in the figure), where the electron and hole states with the opposite spin direction are coupled by the the pairing potential, a gap of $2 \Delta_0$ is opened. A non-trivial superconducting state supporting Majorana zero-modes requires $\Delta_0<\Delta_Z$ (see Ref.~\onlinecite{Weithofer2013}). Under this condition the helical edge states are not gapped at the Fermi energy, which makes the detection of Majorana zero modes more difficult.
As an example, Fig.~\ref{edge_states_dispersion} shows the numerically calculated energy dispersion in a QW without [Fig.~\ref{edge_states_dispersion}a)] and with [Fig.~\ref{edge_states_dispersion}b)] proximity to an s-wave superconductor for $\Delta_0=0.5$~meV$<\Delta_{E}=\Delta_{H}=1.5$~meV. Taking into account that in HgTe QWs the $g$
factor of the $H$ subband is negligibly small in comparison with that of the $E$-subband \cite{Koenig2008} we can set $\Delta_{H}=0$ in our calculations. Moreover, the edge states are composed of mainly the $H$-component which leads only to a small Zeeman splitting (smaller than $\Delta_0$) of the edge states even in the regime of non-trivial superconductivity with $\Delta_0<\Delta_E$. This issue is illustrated in Fig.~\ref{edge_states_dispersion}c) for the case without superconductivity and in Fig.~\ref{edge_states_dispersion}d) with proximity to an s-wave superconductor. In a realistic parameter regime the bulk of the TSC can be non-trivial, and at the same time the helical edge states can be gapped.

For $\Delta_0>\Delta_{E}$ the zero bias conductance of NS junctions is kept at the value of $4e^2/h$ because the S-region lies in the topologically trivial phase without Majorana zero-energy states  \cite{Oreg2010_S, Weithofer2013}. In contrast, for $\Delta_0<\Delta_{E}$ the conductance switches between $4e^2/h$ and $2e^2/h$, depending on the number $N$ of bulk subbands crossing the Fermi level in the S-region in the absence of the pairing potential $\Delta$ (see Fig.~\ref{conduct_invert}). Fig.~\ref{dispersion_N} shows the energy dispersion in the QW without $\Delta$ corresponding to the black points in Fig.~\ref{conduct_invert}a) for different values of $N$. Here, for a fixed value of the Zeeman term $\Delta_E$, $N$ increases with increasing the negative value of the Fermi energy. 
In the presence of the superconducting pairing a gap opens in the spectrum near the Fermi energy, whereby it almost vanishes in the blue regions shown in Fig.~\ref{conduct_invert}b). Thus changing the number of the occupied subbands in the QW in the absence of superconductivity leads to the closing and re-opening of the bulk gap in the presence of superconductivity, and, consequently, to the alternation of the topologically trivial ($Q=1$) and non-trivial ($Q=-1$) superconducting phases in the superconducting QW. The non-trivial character of the superconducting phase can be clearly identified by the quantized conductance value of $2e^2/h$ [see Fig.~\ref{conduct_invert}c)] as well as by the finite probability density at the interface of N- and S-layers [see Fig.~\ref{conduct_invert}d)].
It should be noted that the topological phase in the S region depends also on the width of the QW $W$ as well as on the structure inversion asymmetry, see Fig.~\ref{conduct_QSHI_N}a) and b) where the conductance in a QSHI-TSC junction switches between $4e^2/h$ and $2e^2/h$ depending on the values of $\Delta_E$, $W$ and the Rashba term $\alpha$.

Due to the finite width of the NS structure a zero-energy bound state shows up in the probability density in the TSC at the boundary with the N-region when the width $W=250$~nm [see Fig.~\ref{prob_density}b)]. It should be noted that in this case the superconducting coherence length $\zeta=\hbar v_F/ \Delta_0$, with $v_F$ the Fermi velocity, is comparable to $W$ and we can refer to this case as the 1D TSC regime. However, for the relatively large width of the structure $W=1000$~nm~$\gg\zeta$ we can see a spreading of the MBS along the edges of the superconducting region (see Fig.~\ref{prob_density}c)). Recognizing this, as well as the fact that the range of parameters where the non-trivial superconducting state is realized coincides with the one of the 2D TSC [see Fig.~\ref{conduct_invert}e)] we can identify this case as the 2D TSC regime. Similar behavior of the zero-energy Majorana state has been reported in a quasi-one-dimensional $p_x+i p_y$ superconductor \cite{Potter2010}. 

\onecolumngrid
\begin{center}
\begin{figure}[ht!]
\includegraphics[width=12cm]{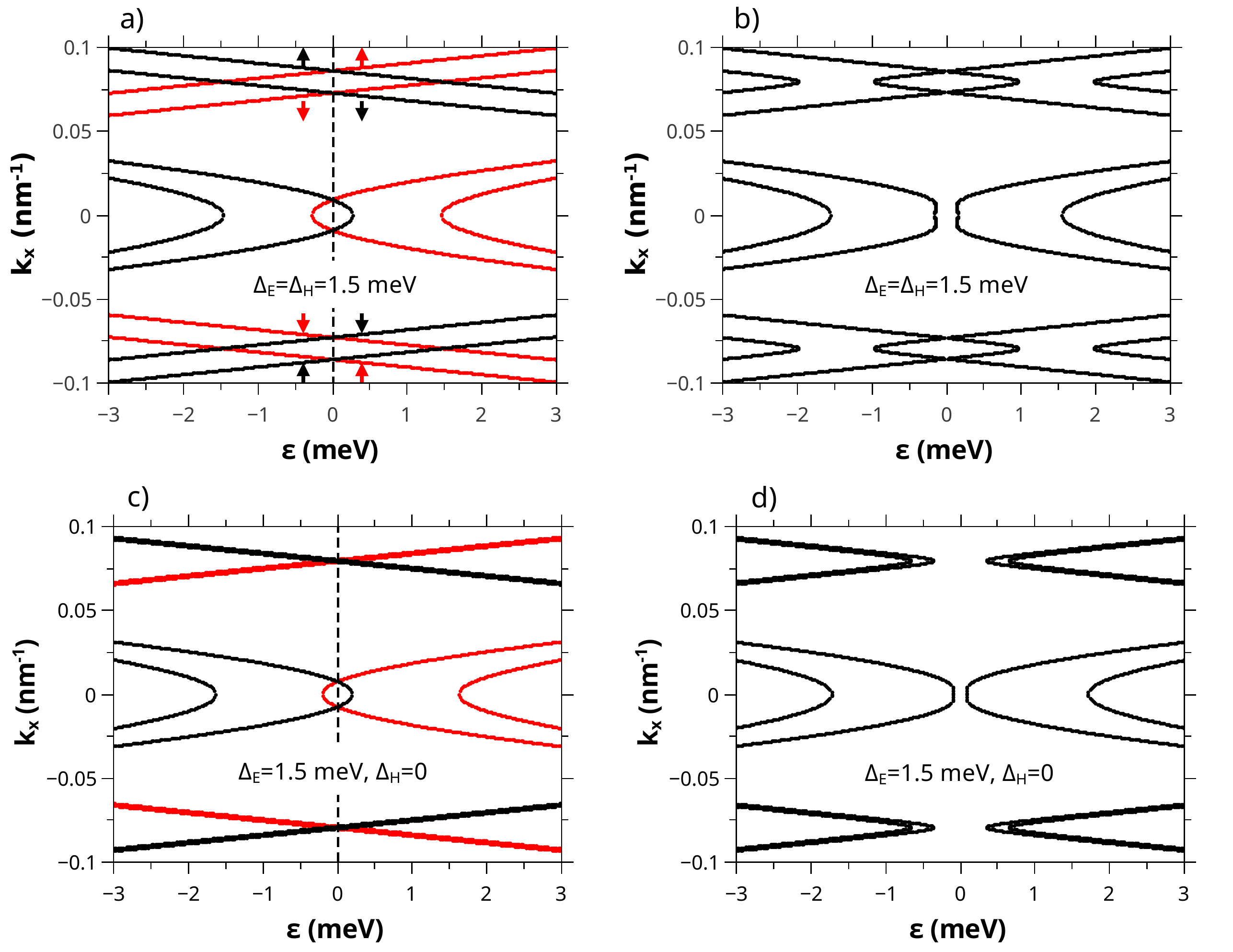}
\caption{Energy dispersion in a HgTe QW without [plots a) and c)] and  with [plots b) and d)] proximity to an s-wave superconductor. The calculations have been done for $M=-10$~meV, $C_{N,S}=9.7$~meV, $W=250$~nm, $E_F=0$, $\alpha=0$. Zeeman splitting $\Delta_{E}=\Delta_{H}=1.5$~meV in a) and b) plots, and $\Delta_{E}=1.5$~meV, $\Delta_{H}=0$ in c) and d) plots. In a) and c) black (red) lines correspond to the electron (hole) energy dispersion, and the arrows represent the spin direction of the corresponding states.}
\label{edge_states_dispersion}
\end{figure}
%
\begin{figure}[ht!]
\includegraphics[width=12cm]{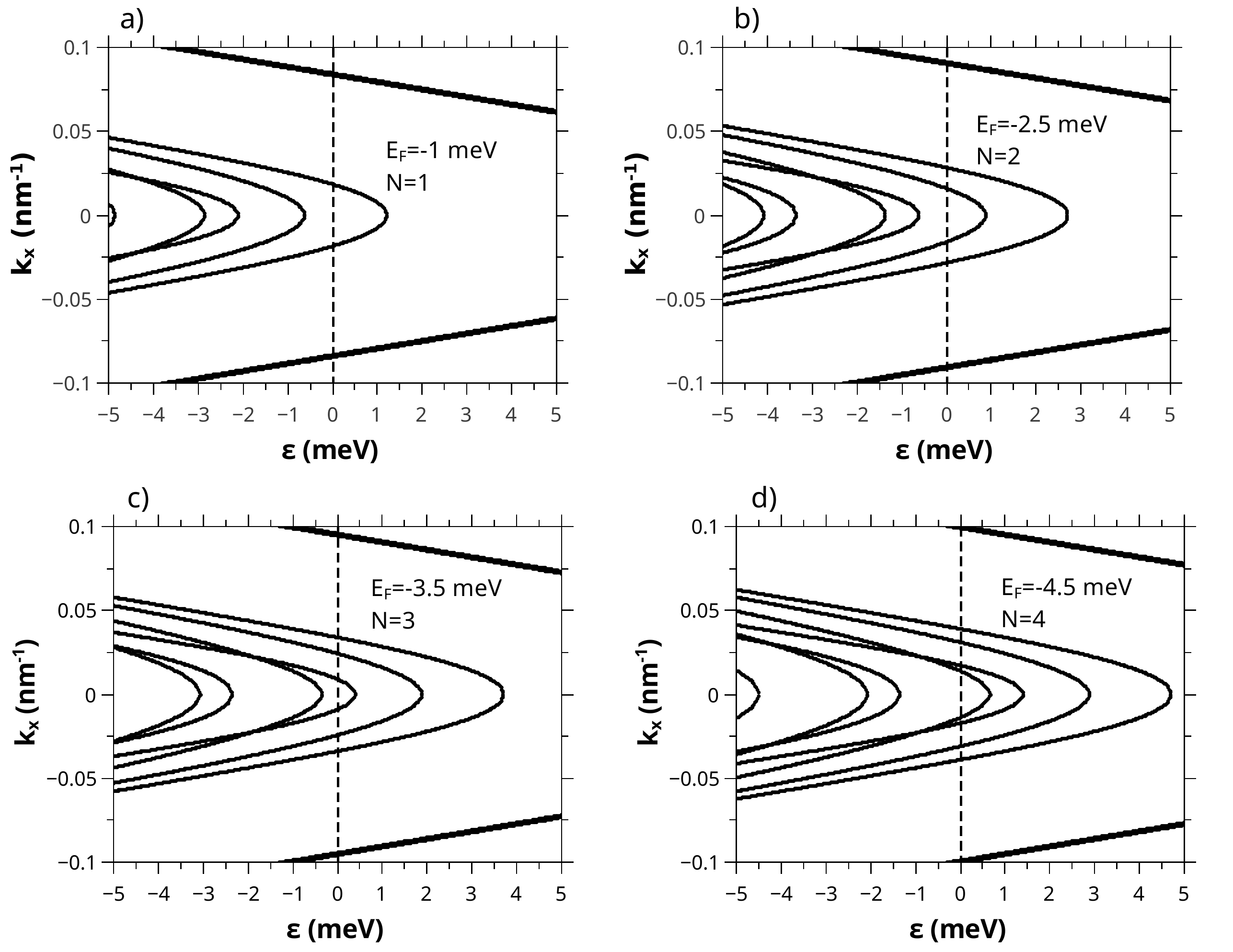}
\caption{Energy dispersion in a HgTe QW with different number $N$ of occupied bulk subbands. The calculations have been performed for $M=-10$~meV, $C_N=9.7$~meV, $W=250$~nm, $\alpha=0$, $\Delta_{E}=1.5$~meV, $\Delta_{H}=0$. The dashed line shows the position of the Fermi level for $\varepsilon=0$.}
\label{dispersion_N}
\end{figure}
\end{center}
\twocolumngrid
\begin{figure}[ht!]
\includegraphics[width=7cm]{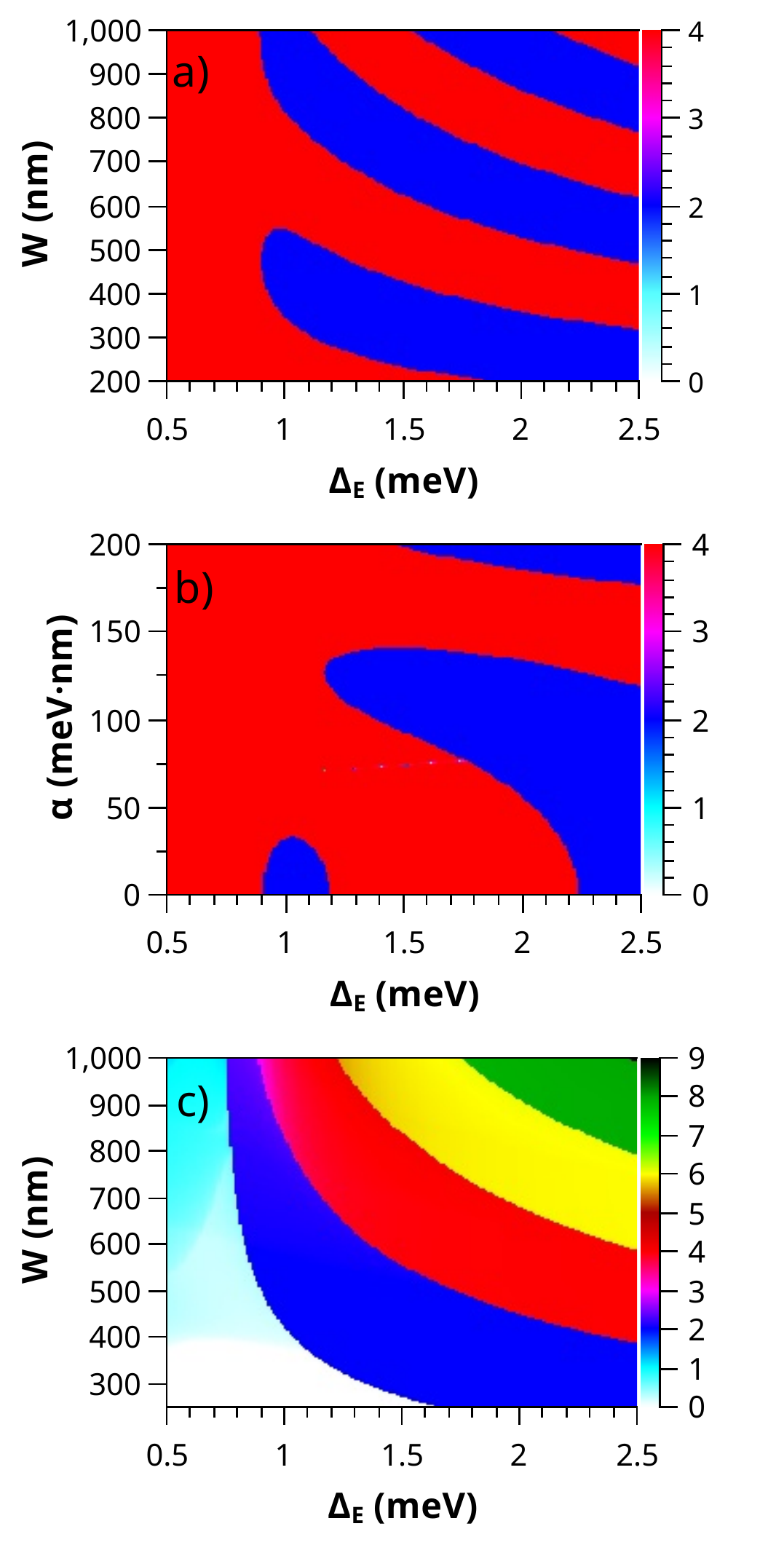}
\caption{Differential conductance (in units of $e^2/h$) in the NS structure based on the QSHI in a) and b) and on the non-inverted QW in c) as a function of Zeeman term $\Delta_{E}$ and width $W$ in a) and c) or Rashba spin-orbit term $\alpha$ in b). The calculations are presented for $\varepsilon=0$ and $E_F=0$. Other parameters are $M=-10$~meV, $C_N=0$, $C_S=9.7$~meV in a) and b) and $M=10$~meV, $C_N=-11$~meV, $C_S=-9.7$~meV in c), $W=500$~nm in b) and $\alpha=0$ in a) and c).}
\label{conduct_QSHI_N}
\end{figure}

\subsection{NS junctions based on non-inverted QWs}
\begin{center}
\begin{figure}[h!]
\includegraphics[width=12.5cm]{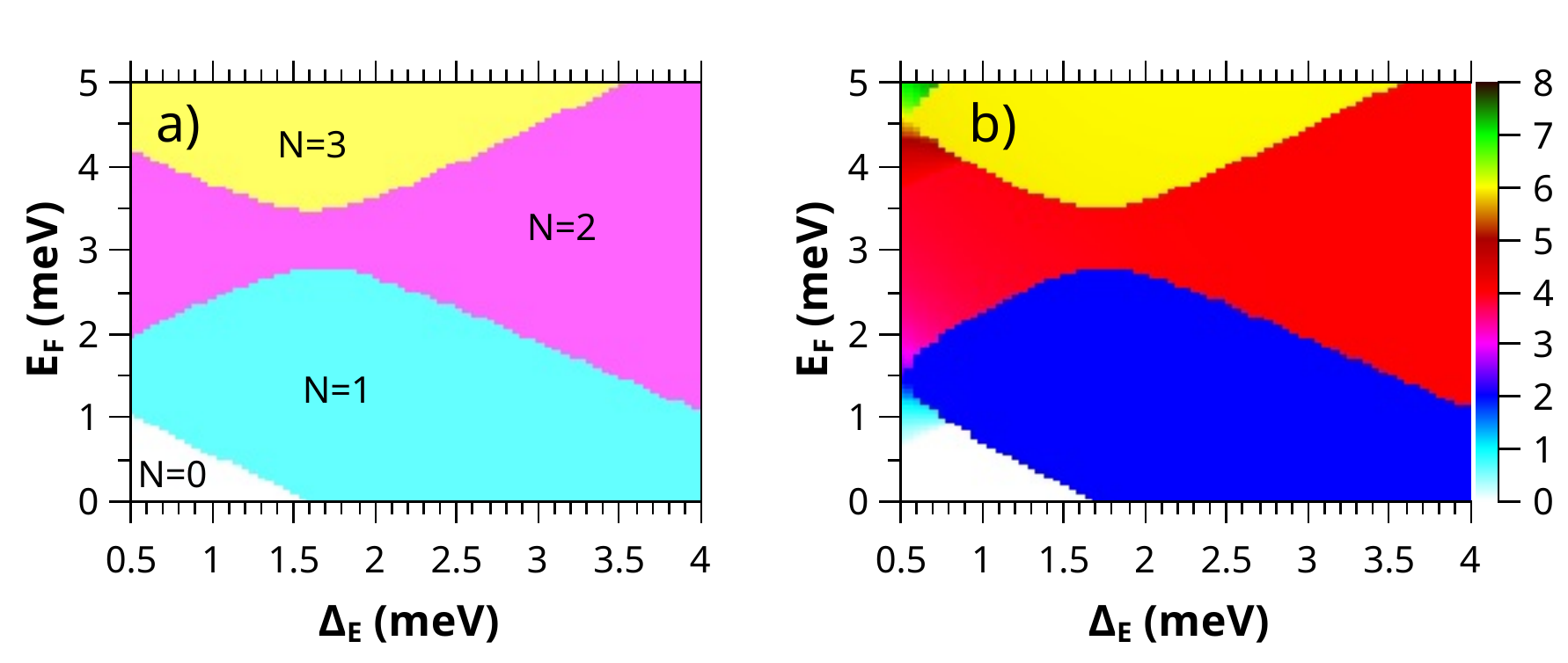}
\caption{a) Diagram representing the number of the occupied subbands in a non-inverted QW at the Fermi level as a function of the Zeeman term and Fermi energy. b) Conductance (in units of $e^2/h$) in the NS structure based on a non-inverted QW. All plots are presented as a function of the Zeeman term $\Delta_{E}$ and the Fermi energy $E_F$ for the QW with a width $W=250$~nm, $M=10$~meV, $\varepsilon=0$, $\alpha=0$. The doping parameter is $C_N=-9.7$~meV in a), $C_N=-11$~meV, $C_S=-9.7$~meV in b).}
\label{conduct_noninvert}
\end{figure}
\end{center}

Next, we consider NS junctions based on non-inverted QWs (since non-trivial TSC phases are still possible in these QWs) and compare these structures with the QSHI-S junctions. In analogy with the inverted QW regime the conductance of the NS structure depends on the number $N$ of bulk subbands crossing the Fermi level in the S-region in the absence of the pairing potential (see Fig.~\ref{conduct_noninvert}). In contrast to the QSHI-S junction, the conductance takes quantized values $G=(2e^2/h)n$, where $n$ is not restricted to be 1 or 2 but takes odd (even) values for the topologically non-trivial (trivial) S phase with $Q=-1$ ($Q=1$). As it can be seen from Fig.~\ref{conduct_noninvert}b), $n=0,1,2,3$ for the white, blue, red and yellow regions in the plot. However, conductance quantization in the non-inverted structures is not protected because of imperfect Andreev reflection of bulk states at the NS interface [see Fig.~\ref{conduct_QSHI_N}c)]. An alternation of the trivial and non-trivial topological superconducting phases has been reported also in spin-orbit-coupled multichannel semiconducting nanowires in proximity to an s-wave superconductor, where protected Majorana modes are predicted to appear at the ends of the wire with an odd number of channels, whereas an even number of the occupied subbands corresponds to the trivial superconducting phase \cite{Lutchyn2011,Wimmer2011,Stanescu2011_S,Reuther2013}. But from an experimental point of view it could be difficult to determine the number of occupied subbands in the system, similar to the case of an NS junction based on a non-inverted QW, and, as a consequence, correctly identify the topological character of the superconductor.

We note in passing that the propagating states of the chiral Majorana edge mode above the minigap (see Fig.~\ref{edge_states_dispersion}d) and Fig.~\ref{scheme_NS}c)), can be probed in transport at finite bias voltage $V$ in the non-inverted regime (or in the inverted regime, when the Fermi level is above the gap of the QSH insulator) leading to rather common {\it non-quantized} conductance values.
\onecolumngrid

\FloatBarrier
\twocolumngrid


\begin{thebibliography}{99}
\bibitem{basis} The field operators in $H_{{\rm NS}}$ are then given in the corresponding Nambu basis $\Psi({\bm r})=(\psi_{E+}({\bm r}),\psi_{H+}({\bm r}),\psi_{E-}({\bm r}),\psi_{H-}({\bm r}),\psi_{E-}^{\dagger}({\bm r}),\psi_{H-}^{\dagger}({\bm r}),\\-\psi_{E+}^{\dagger}({\bm r}),-\psi_{H+}^{\dagger}({\bm r}))^{T}$.
\bibitem{param3} We choose in our calculations $\Delta_0=0.5$~meV, $\phi=0$.
\end{thebibliography}
\end{document}